\documentstyle{article}
\title{\LARGE Conformal quantum space $SU(2,2)\simeq O(2,4)$ with Poincare group of motion}
\author{A.~N.~Leznov\thanks{ Universidad Autonoma del Estado de Morelos, CCICAp,Cuernavaca, Mexico}} \date{}

\newcommand{\rig}[2]{\stackrel{#2\rightarrow}{#1}}

\begin{document}
\maketitle

\maketitle

\begin{abstract}
It is shown that algebra of quantum space of the title of the present paper may be realized on usual unphysical Minkowskii one. Equations of field theory and there solutions are discussed. Solution equations of particle motion are obtained in implicit form from the classical ones.   
\end{abstract}

\section{Introduction}

In \cite{LEZM} it was proposed the way for construction gauge-invariant field theory in quantum space with 3 dimensional constants. 
In the case of the quantum space with infinite value of the constant with dimension of the length this construction become essentially more simple. And thus it is the sense to consider this version in the first turn before over going to the general case.

In this version group of inner symmetry coincide with Poincare one and thus equation of field theory will conserve its form in terms of operators of space-time translations $p_i$. But the form of this operators will be defined on the space of representation of corresponding quantum algebra (see (\ref{2}) in the next section). In the usual Minkovski space-time $p_i=\frac {\partial}{\partial x_i}$. As it was shown in \cite{LEZM} in the case under consideration quantum algebra can be realized on two unit four-dimensional and two-dimensional vectors and corresponding formulae for all elements of quantum space where presented there in explicit form. In the present paper we essentially simplify
form of these elements showing that realization in \cite{LEZM} is canonically equivalent to realization on unphysical Minkovski space. In other words we present realization
quantum space on unphysical four dimensional manifolds with coordinates $x_i^M$ with commutation relations with physical space-time translations $[p_i, x_j^M]=g_{i,j}$.
All other generators of quantum space are  presented in terms of these conjugated pairs of Darboux  \cite{ESEN}. For the reader for whom mathematical problem not so essential author advise begin reading directly from section 5.

\section{Quantum ("deformed") spaces}

The most general form of the commutation relations between the elements of the quantum  four-dimensional space-time ($x$- coordinates, $p$- impulses, $F$- generators of Lorentz rotations, $I$ -unity element) are the following ones \cite{KHL}
$$
[p_i,x_j]=ih(g_{ij}I+{F_{ij}\over H}),\quad  [p_i,p_j]={ih\over L^2}F_{ij},
\quad [x_i,x_j]={ih\over M^2}F_{ij},
$$
\begin{equation}
[I,p_i]=ih({p_i\over H}-{x_i\over L^2}), \quad [I,x_i]=ih({p_i\over M^2}-
{x_i\over H}),\quad [I,F_{ij}]=0 \label{2}
\end{equation}
$$
[F_{ij},x_s]=ih(g_{is}x_j-g_{js}x_i),\quad [F_{ij},p_s]=ih(g_{is}p_j-g_{js}p_i)
$$
$$
[F_{ij},F_{sk}]=ih(g_{js}F_{ik}-g_{is}F_{jk}-g_{jk}F_{is}+g_{ik}F_{js})
$$
Commutation relations (\ref{2}) must be submitted by some additional conditions which responsible for correct limit to usual Minkowski space-time in the infinite limit of dimensional parameters. Such conditions looks as
$$
IF_{i,j}={x_j p_i-x_i p_j+p_i x_j-p_j x_i \over 2}
$$
which in Minkovski limit ($I\to 1$) represent relation between angular moments and linear coordinates and impulses.
The additional conditions above allow to choose definite representation of algebra 
(\ref{2}) in the unique way \cite{LEZM}.

Commutation relations of the quantum space contain 3 dimensional parameters
of the dimension length $L$, the impulse $Mc\to M$ and the action $H$. The
equalities of Jacobi are satisfied for (\ref{2}). It should be stressed the signs 
of $L^2,M^2$ are not required to be positive.
The limiting procedure $ M^2,H\to \infty$ leads to the space of
constant curvature, considered in connection with Column problem by
E.Schredinger \cite{SCH}, $L^2,H\to \infty$ leads to quantum space of Snyder
\cite{I}, $H\to \infty$ leads to Yangs quantum space \cite{Y}. Except of
$L^2,M^2$ parameter dimension of action $H$ was introduced in  \cite{KHL}.

The term quantum  space is not very useful, because the modified classical dynamics may be considered in it (also as electrodynamics, gravity theory and so on).

In general (\ref{2}) is commutation relations one of real forms of six-dimensional rotation group. In the case $L^2\to \infty$ as it follows directly from (\ref{2}) commutation relations between $p_i,F_{i,j}$ exactly coincide with commutation relation of Poincare algebra and thus the last is the group of motion of the space-time under consideration. But this is not the Snyder case. Because in the case under consideration the whole algebra (\ref{2}) coincide with conformal one $SU(2,2)\simeq O(2,4)$. In the Snyder case ($H\to \infty$) this is algebra of 5-dimensional rotation with shifts. This is not a semi simple algebra.

\section{Realization of quantum ("deformed") space}

Let us introduce in (\ref{2}) dimensional less variables ( $0\leq M^2$)
$$
x_{\alpha}={h\over M}O_{4,\alpha}, \quad x_0={h\over M}O_{4,5}, p_{\alpha}={h\over H}M (O_{4,\alpha}+O_{6,\alpha}), \quad p_0={h\over H} M (O_{4,5}+O_{6,5})
$$
\begin{equation}
I={h\over H}O_{6,4}, \quad f_{\alpha}=h O_{5,\alpha}, \quad l_{\alpha,\beta}=h O_{\alpha,\beta}
\label{DEF}
\end{equation}
In formulae above indexes $\alpha,4$ belong to compact group of four dimensional rotation
$O(4)$, $5,6$ to compact group $O(2)$. $O_{í,j}$ generators of non compact algebra $SU(2,2)=O(2,4)$.

The second Kazimir operator in this notations looks as
$$
K_2=I^2+{(\rig{p}{})^2-p_0^2\over M^2}-{(\rig{p}{}\rig{x}{})+(\rig{x}{}\rig{p}{})-p_0x_0-x_0p_0\over H}+{(\rig{f}{})^2-(\rig{l}{})^2\over H^2}= 
$$
$$
({h\over H})^2(O_{6,4}^2+\sum O_{6,\alpha}^2+O_{5,4}^2+\sum O_{5,\alpha}^2-O_{6,5}^2-\sum O_{\alpha,\beta}^2)
$$

In \cite{LEZM} it was shown that to have correct limit to usual Minkovski space it is necessary use only definite unitary representation of six dimensional non compact rotation group. Generators of this representation must satisfy 15 additional conditions namely
$$
\sum \epsilon_{i,j,k,l,m,n} O_{k,l} O_{m,n}=0
$$
where $\epsilon_{i,j,k,l,m,n}$ anti symmetrical tensor Levy-Chevita in six dimensional space.
In cited papers such representation was found and present below generators of conformal algebra in this representation
$$
O_{6,4}=\cos \phi(\rho q^1_4+\sum q^i_4 Q^{i,1})-\sin \phi q^1_4 \frac{\partial}{\partial \phi}
$$
$$
O_{5,4}=\sin \phi (\rho q^1_4+\sum q^i_4 Q^{i,1})+\cos \phi q^1_4 \frac{\partial}{\partial \phi}
$$
\begin{equation}
O_{6,\alpha}=\cos \phi(\rho q^1_{\alpha}+\phi \sum q^i_{\alpha} Q^{i,1})-\sin \phi q^1_{\alpha} \frac{\partial}{\partial \phi} \label{RAC}
\end{equation}
$$
O_{5,\alpha}=\sin \phi(\rho q^1_{\alpha}+\sum q^i_{\alpha} Q^{i,1})+\cos \phi q^1_{\alpha}\frac{\partial}{\partial \phi}
$$
$$
O_{6,5}=\frac{\partial}{\partial \phi}, \quad O_{\alpha,\beta}=Q_{\alpha,\beta}
$$
where $Q_{\alpha,\beta}, \tilde Q^{\alpha,\beta}$ generators of left and right regular representation of compact group of four dimensional rotations, $q^{\alpha}_{\beta}$ elements of four dimensional orthogonal matrix $q=\exp O_{1,2}\psi \exp O_{2,3}\theta \exp O_{3,4}\sigma$. In (\ref{RAC}) it is necessary work in invariant subspace which is defined by conditions $\tilde Q^{1,2}=\tilde Q^{2,3}=\tilde Q^{1,3}=0$,which means invariance with respect to transformations of three dimensional right group of rotations.  $\rho$ parameter defining representation of conformal algebra. In its terms second Kazimir operator looks as $K_2=({h\over H})^2\rho(\rho+4)$.

Directly from (\ref{RAC}) it follows immediately that non compact generators will be anti hermitian under the choice $\rho=-2+i\delta$, compact generators are anti hermitian ones $Q^H=-Q$. Thus if we want to work with hermitian operators of observable variables all generators above its necessary multiply on i. In this case Kazimir operator of main continues series
take the form $K_2={h\over H})^2(4+\delta^2)$ and allow correct  (by sign) limit to Minkowsky space-time.

\section{Rewriting results of the previous section in more observable form}

In this section we realize representation of the quantum space on the variables usual to Minkovski space four coordinates and four conjugated for them impulses. But additional coordinates do not coincide with physically observable ones.

From the point of view of functional algebras \cite{ESEN} this correspond to description on the language of Darboux pairs of conjugated coordinates and impulses.
    
It is necessary keep in mind (as it was mentioned above) that representation (\ref{RAC})
is realized on four dimensional unit vector $q^1$ and unit two dimensional vector $\phi$. And thus functions on this space of representation are dependent on four independent coordinates. We will use free first components of four dimensional unit vector $q$ and $\phi$. In other words arbitrary function on the space of representation has the form $F=F(q^1_{\alpha},\phi)$.
Dimensionless operators of impulses (space-time translations) have the form (\ref{DEF}) and (\ref{RAC})
$$
p_0=\rho \sin \phi q^1_4+(1-\cos \phi q^1_4)\frac{\partial}{\partial \phi}+\sin \phi q^1_4 (\rig{q}{}\frac{\partial}{\partial \rig{q}{}})
$$
In writing this expression we take into account that action $\tilde Q^{i,1}F(q^1_{\alpha},\phi)=
\sum_{\alpha} [\tilde Q^{i,1},q^1_{\alpha}]\frac{\partial F}{\partial q^1_{\alpha}}=\sum_{\alpha}q^i_{\alpha}\frac{\partial}{\partial q^1_{\alpha}}F$ and the same equalities in corresponding calculations below.
$$
p_{\alpha}=\rho\cos \phi q^1_{\alpha}+(q^1_4-\cos \phi)\frac{\partial}{\partial q^1_{\alpha}} +q^1_{\alpha}\cos \phi(\rig{q}{}\frac{\partial}{\partial \rig{q}{}})+q^1_{\alpha}\sin \phi\frac{\partial}{\partial \phi}
$$
where $(\rig{q}{}\rig{q}{})+(q^1_4)^2=1$.
By direct simple computation one can verify that all operators above are mutual commutative. 

Now we resolve the last system of equations with respect to operators  $\frac{\partial}{\partial q^1_{\alpha}},\frac{\partial}{\partial \phi}$

Multiplying equations for $p_{\alpha}$ on  $q^1_{\alpha}$ and perform summation we come to linear system of algebraic equations for $X=\rig{q}{}\frac{\partial}{\partial \rig{q}{}}),Y=\frac{\partial}{\partial \phi}$
$$
\sin \phi q^1_4 X+(1-q^1_4\cos \phi)Y=p_0-\rho\sin \phi q^1_4,
$$
$$
q^1_4 (1-q^1_4\cos \phi)X+\sin \phi(1-(q^1_4)^2)Y=\sum (q^1_{\alpha}p_{\alpha})-\rho\cos \phi (1-(q^1_4)^2)
$$
Resolving this system of equations leads to derivatives
$$
\frac{\partial}{\partial \phi}=-{\rho \sin \phi\over q^1_4-\cos \phi}+{(1-q^1_4\cos \phi)p_0-\sin \phi \sum (q^1_{\alpha}p_{\alpha})\over (q^1_4-\cos \phi)^2}
$$
$$
\frac{\partial}{\partial q^1_{\alpha}}=-{\rho q^1_{\alpha}\over q^1_4(q^1_4-\cos \phi)}+[{p_{\alpha}\over (q^1_4-\cos \phi)}+{-p_0\sin \phi+\sum (q^1_{\alpha}p_{\alpha})\over q^1_4(q^1_4-\cos \phi)^2}]
$$

Using the same technique of calculations it is not difficult to check that four coordinates
(dimensionless)
$$
\bar x_0=-{\sin \phi\over q^1_4-\cos \phi},\quad \bar x_{\alpha}={q^1_{\alpha}\over q^1_4-\cos \phi}  
$$
and dimensionless impulses $p_{\alpha}=(O_{4,\alpha}+O_{6,\alpha}), p_0=(O_{4,5}+O_{6,5})$ satisfy usual canonical commutation relations
$$
[p_i,x_j]=g_{i,j},\quad g_{0,0}=-1,\quad g_{\alpha ,\beta}=1 
$$
These coordinates are not physical observable coordinates of the particles but have only the mathematical meaning as additional way for enumerating points in quantum space in representation canonically equivalent to the initial one (\ref{DEF}) and (\ref{RAC}). We will marked these coordinate by additional $M$ upper indexes.  Physically observable coordinates are defined by (\ref{DEF}) and may be expressed in terms of conjugated pairs $p_i,x_i$. Indeed
$$
x_0=O_{4,5}=p_0-O_{6,5}=p_0-\frac{\partial}{\partial \phi}=\rho x_0^M+{q^1_4\over q^1_4-\cos \phi}p_0-x_0^M(x_0^M p_0-(\rig{x}{M},\rig{p}{})=
$$
$$
\rho x_0^M+[{(x_0^M)^2-(\rig{x}{M})^2+1)\over 2}p_0-x_0^M(x_0^M p_0-(\rig{x}{M},\rig{p}{})]
$$ 
In the last transformation we have used equality 
$$
(x_0^M)^2-(\rig{x}{M})^2={(\sin \phi)^2-(1-(q^1_4)^2)\over (q^1_4-\cos \phi)^2}={q^1_4+\cos \phi\over q^1_4-\cos \phi}
$$  

Further
$$
x_{\alpha}F=O_{4,\alpha}F=\sum_{\beta} [Q_{4,\alpha},q^1_{\beta}]\frac{\partial F}{\partial q^1_{\beta}}=-q^1_4 \frac{\partial }{\partial q^1_{\alpha}} F
$$
Now substituting equated above $\frac{\partial }{\partial q^1_{\alpha}}$ we come to the expression for space physical coordinate
$$ 
x_{\alpha}=\rho x^M_{\alpha}+[{(x_0^M)^2-(\rig{x}{M})^2+1)\over 2}p_{\alpha}-x^M_{\alpha}(x_0^M p_0-(\rig{x}{M},\rig{p}{})
$$

\section{Realization of quantum space on the usual Minkovski one}

It is possible present all generators of the conformal algebra of quantum space in terms of Darboux pairs of generators of Minkovski space $p_i,x^M_j$. For dimensionless generators we have
$$
x_i=\rho x^M_i+[{x_M^2\over 2}p_i-x^M_i(x^M p)]
$$
where $x_M^2=(x_0^M)^2-(\rig{x}{M})^2,(x^M p)=x_0^M p_0-(\rig{x}{M},\rig{p}{})$.
$$
I=\rho-(x^M p),\quad F_{j i}=(x_j^M p_i-x_i^M p_j)
$$
The presented above generators satisfy commutation relations of quantum algebra 
(\ref{2}) with $L=\infty$ the satisfy additional conditions
$$
IF_{i,j}={x_j p_i-x_i p_j+p_i x_j-p_j x_i \over 2}
$$
These conditions are responsible for correct relations between angular and linear coordinates and impulses. 

\section{Field theory and equations of classical dynamics in quantum space}

All equations of field theory contain its form in quantum space with Poincare group of motion. Only one condition was used in construction of field theory in Minkovski space. Namely condition of invariance of its equations with respect transformation of Poincare group. Thus in all equations of the classical field theory it is necessary change operators of differentiation on space-time coordinates on corresponding operator of translation in quantum space. If we will enumerate quantum space by un physically coordinates $x^M_i$ the solution of the field equations will coincides with the classical ones. But if it will be necessary to solve some problem with initial conditions in real physical space time it will be necessary enumerate the points by proper values of some commutative four generators having physical sense. For instance in the space under consideration impossible enumerate quantum space by proper values of four
coordinates there generators are not commutative and have no common proper values. But four generators $t,(\rig{r}{})^2,(\rig{l}{})^2,l_3$ have common proper values. Equations of field theory it is possible to write and solve in these coordinates. Of course there form and solution will be different from classical ones.

We would like clarify this situation on the example of solution of the Kepler problem of classical dynamics. The equation of Hamilton-Yacobi conserve its form
$$
(P_0S-A_0)^2-((\rig{p}{}S-(\rig{A}{})^2=m^2
$$
where $A$ vector potential of electro magnetic field. Solution equations of Makswell in unphysical coordinates is the same and particulary coincide with usual potential of Column. And thus in unphysical coordinates we will have solution of classical Kepler problem with conserving Laplace vector in non relativistic limit.
But this solution must be substituted into equations connected unphysical
coordinates and impulses with the physically ones by the formulae of the previous section. Then the parameter of the proper time, which parametrize
solution in relativistic case will be connected with the real time coordinate and all the relation of connection space coordinates will give modified traectory of motion in  quantum space under consideration. Of course from these results it will be possible to give rough estimation for parameters $H,M$ of deformed space-time.

\section{Outlook}

In present paper we have shown how necessary representation of quantum space algebra may obtain in usual for physics Minkovski space. Of course this is equivalent to obtained initilly in
\cite{LEZM}. But construction of the present paper allow to work in quantum space practically
by the same way as in the Minkovski of the usual theory of the field. From results of the paper it became clear that in the general case of quantum space with 3 dimension parameters its algebra may be realized on un physical Mincovski space. It follows from application of Darboux theorem  \cite{ESEN} to quantum space algebra of the general position. Thus all physical consequence author will try to do after consideration of the general case which he hope to finish in the nearest time. 

\section{Acknowledgments}

Author thanks CONNECUT for finance support.


\begin{thebibliography}{9}

\bibitem{LEZM} A.N. Leznov {\it Quantized spaces are four-dimensional compact manifolds with de-Sitter (O(1,4) or O(2,3)) group of motion.e-Print: hep-th/0410255 }
{\it  Theory of fields in quantized spaces. e-Print: hep-th/0409102}
\bibitem{ESEN} L.P.Eisenhart {\it Continuous Groups of Transformations, Princiton N.Y.,
Princiton University Press 1933}  
\bibitem{SCH} Schrodinger E  {\it Proc.R.Irish.Acad. A 46, 9 (1940)}
\bibitem{I} Snyder {\it Phys.Rev. 71 (1947) 38}, Ya.Golfand JETP 37 504 (1959), V.G.Kadyshevsky JETP 41 1885 (1961)
\bibitem{Y} Yang {\it Phys.Rev. 72 (1947) 874}
\bibitem{KHL} A.N.Leznov and V.V.Khrushov {\it Preprint INEP 73-99 (1973), Grav.Cosmol.9:159,2003. e-Print: hep-th/0207082}
\bibitem{LEZQ} A.N. Leznov {\it Nuclear Physics B640 [PM] (2002) 469-480}
\bibitem{LM} A.N. Leznov and J.Mostovoy {\it Classical dynamics in deformed spaces. e-Print: hep-th/0208152 Published in J.Phys.A36:1439-1450,2003}.
\end{thebibliography}
\end{document}